\documentclass{pasj00}

\SetRunningHead{S. Konami et al.}{
Metallicity in the ISM of NGC 4258}

\title{Suzaku Observation of the Metallicity \\in the Interstellar Medium of NGC 4258}
\author{
 Saori \textsc{Konami},\altaffilmark{1,2}
 Kosuke \textsc{Sato},\altaffilmark{3} 
 Kyoko \textsc{Matsushita},\altaffilmark{1} 
 Shin'ya \textsc{Yamada},\altaffilmark{4}
 Naoki \textsc{Isobe},\altaffilmark{5,2}\\
 Atsushi \textsc{Senda},\altaffilmark{6,2} 
 Asami \textsc{Hayato},\altaffilmark{1,2} 
 Poshak \textsc{Gandhi},\altaffilmark{2}
 Toru \textsc{Tamagawa},\altaffilmark{2,1}
and Kazuo \textsc{Makishima},\altaffilmark{4,2}
}
\altaffiltext{1}{
Department of Physics, Tokyo University of Science,
1-3 Kagurazaka, Shinjuku-ku, Tokyo 162-8601}
\altaffiltext{2}{
Cosmic Radiation Laboratory, the Institute of Physical and Chemical Research,
2-1 Hirosawa, Wako, Saitama 351-0198}
\email{konami@crab.riken.jp}
\altaffiltext{3}{
Graduate School of Natural Science and Technology,
Kanazawa University, 
Kakuma, Kanazawa, Ishikawa 920-1192}
\altaffiltext{4}{
Department of Physics, University of Tokyo,
7-3-1, Hongo, Bunkyo-ku, Tokyo, 113-0033}
\altaffiltext{5}{
Department of Astronomy,Kyoto University,
Kitashirakawa-Oiwake-cho, Sakyo-ku, Kyoto, 606-8502}
\altaffiltext{6}{
Japan Science and Technology Agency,
Kawaguchi Center Building, 4-1-8, Honcho, Kawaguchi-shi, Saitama 332-0012}
\Received{}
\Accepted{}
\Published{---}

\begin{document}
\maketitle
\begin{abstract}
The Suzaku X-ray satellite observed the nearby spiral galaxy NGC 4258 for 
a total good exposure time of 100 ks. 
We present an analysis of the Suzaku XIS data, in which we 
confirm that the 0.5--2 keV spectra of 
the interstellar medium (ISM) are 
well-represented by a two-temperature model.  The cool and hot ISM 
temperatures are 0.23$^{+0.01}_{-0.02}$ and 0.59 $\pm0.01$ keV, 
respectively. Suzaku's excellent spectral sensitivity enables us to 
measure the metal abundances of O, Ne, Mg, Si 
and Fe of the ISM for the first time. 
The resultant abundance pattern of O, Mg, Si, and Fe 
is consistent with that of the new solar abundance table of 
\citet{lodders_03}, rather than \citet{anders_89}.
This suggests that the metal enrichment processes of NGC 4258 and 
of our Galaxy are similar.
\end{abstract}

\section{Introduction}

Metal abundances in the hot X-ray emitting interstellar medium (ISM) 
are important for understanding the star formation history and 
evolution of galaxies. A large fraction of metals in the ISM are 
synthesized by type Ia and type II supernovae (SNe Ia/II). 
O and Mg are predominantly synthesized by SNe II, while Fe is mainly 
produced by SNe Ia. Therefore, the abundance ratios provide useful 
information on the contribution of both types of SNe enriching the
metals. 

Compared to starburst galaxies, normal spiral galaxies have lower 
star formation rates (SFRs), and have lower X-ray surface brightness, 
because the integrated soft X-ray luminosities of spiral galaxies
correlate with  SFR \citep{tullmann_06}.
In addition, these galaxies sometimes exhibit strong X-ray emission 
from the central active galactic nuclei. As a result, it has remained 
rather difficult to constrain the abundances of their hot ISM over a 
wide range of species from O to Fe. 

In order to determine  the metal abundances of O, Ne, S, and Ar from 
the emission lines in H\emissiontype{II} regions, several optical 
observations have been performed (e.g., \cite{perez_07}, \cite{crockett_06}). 
In particular, O emission lines, a good tracer of SNe II, are the 
brightest in the optical band. However, determination of the 
metal abundances, including O, has large uncertainty 
due to dust grains, the composition of which is difficult to ascertain. 
In addition, no significant Fe lines, a good tracer of SN Ia, 
exist in the optical band. 

The Einstein satellite first observed the thermal X-ray emission 
from spiral galaxies \citep{fabbiano_89}, and ROSAT and ASCA subsequently 
measured the temperature and Fe ISM abundances 
\citep{roberts_01,petre_94}. Chandra and XMM-Newton have enabled us to 
study the spatial distribution of metals by utilizing their large 
effective area and high angular resolution \citep{schlegel_03,yang_07}.
However, the O and Mg abundance measurements  are 
difficult due to the intrinsic instrumental background and 
asymmetric energy response in the low energy band below $\sim1$ keV\@.
Because Suzaku XIS \citep{koyama_07} has higher spectral sensitivity 
below $\sim1$ keV, and a lower and more stable background level compared 
to Chandra ACIS and XMM-Newton EPIC, the accuracy of the determination
of O and Mg abundances is improved.
Recent Suzaku observations have revealed the abundance profiles of the ISM 
in X-ray bright elliptical and starburst galaxies in detail 
\citep{matsushita_07, tsuru_07,tawara_08,yamasaki_08,hayashi_09}.

\begin{figure*}
\FigureFile(\textwidth,\textwidth){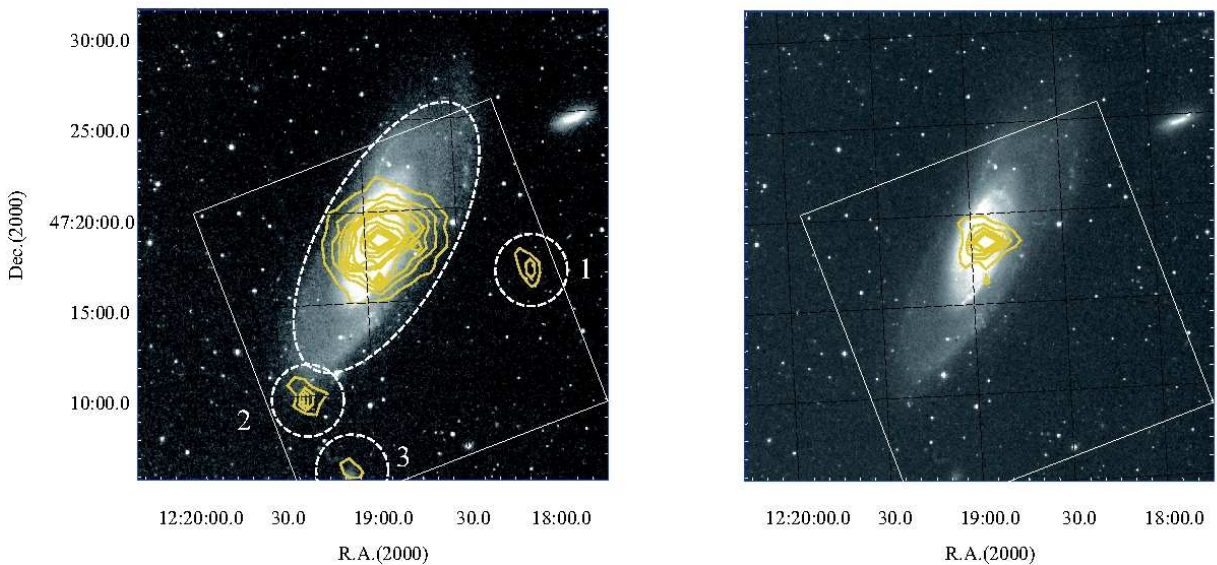}
\caption{The X-ray contour map (orange) in linear scale from
0.5--2 keV (left) and 2--10 keV (right), overlaid on an optical image 
taken by DSS. 
The contour scales range linear by (0.17 -- 5)$\times10^{-2}$ 
and 0.3 -- 1.3 counts sec$^{-1}$ arcmin$^{-2}$, respectively. 
The observed XIS0, 1, 2, and 3 images were added 
on the sky coordinate, and smoothed with $\sigma=12$ pixel 
Gaussian profile. Estimated contributions of the cosmic X-ray background 
(CXB) and the instrumental non X-ray background (NXB) were subtracted, 
although vignetting was not corrected. 
The white square corresponds to the field of view of the Suzaku XIS.
The region where energy spectra were extracted is indicated by a 
white dashed ellipse. Three point sources (white dashed circles) 
were excluded from the background energy spectra.}
\label{image}
\end{figure*}
NGC 4258 (M 106) is a nearby SABbc spiral galaxy with a high inclination 
angle of 72$^\circ$ \citep{tully_88}. Its distance is accurately measured 
to be 7.2 Mpc \citep{herrnstein_99}, where 1$'$ corresponds to 
2.1 kpc. 
\citet{makishima_94} first discovered the 
obscured low-luminosity active galactic nucleus (LLAGN) 
in NGC 4258 with ASCA, and also showed that its 
1--10 keV spectrum is composed of contributions from thermal 
plasmas, integrated low mass X-ray 
binaries (LMXBs), and the nucleus. 
A ROSAT observation detected two thermal components with $kT\sim0.2$ and 
$\sim0.5$ keV in the ISM \citep{vogler_99}, which were
confirmed with Chandra and XMM-Newton observations \citep{yang_07}\@.
The 2--10 keV flux from the nucleus was reported to vary on time scales 
of hours to years \citep{reynolds_00,terashima_02,fruscione_05}.
A recent Suzaku observation has revealed that the circum-nuclear matter 
forms a slim obscuring structure with solid angle much smaller than 
typical thick tori \citep{yamada_09,reynolds_08}\@.   
The galaxy has a curious jet-like feature, what is called an ``anomalous arms'', 
emitting soft X-rays (e.g., \cite{cecil_92},\cite{wilkes_95,cecil_95}). 
In addition, the hot spots associated with the galactic radio jet have been detected 
at a distance of around 1$'$ 
from the nucleus in X-rays by Chandra \citep{yang_07}.
Aside from these anomalous arms, which seem to be related to the nuclear
activity, NGC 4258 can 
be regarded as being similar to our Galaxy in terms of 
the low activity of the central black hole. The fact that the emission from 
the nucleus is deeply absorbed with a thick circum-nuclear matter 
\citep{makishima_94} is helpful to study the thermal emission. 
In this paper, we report the detailed analysis of NGC 4258 with Suzaku,
focusing on the thermal emission of the ISM to constrain the metal
abundances. 

Unless noted otherwise, the solar abundance table used is taken from 
given by \citet{anders_89}, and the quoted errors are for a 90$\%$ confidence 
interval for a single interesting parameter.

\section{Observation and Data Reduction}\label{sec:obs}
Suzaku carried out an observation of NGC 4258 in 2006 June.
We analyzed only the XIS data
in this paper, although Suzaku observed the object with both XIS 
and HXD \citep{takahashi_07}, centered on the HXD field of view, 
and the HXD results from this observation on the nucleus have 
already been published by \citet{yamada_09}. 
The XIS consists of four sets of X-ray CCD camera system (XIS0, 1, 2, 
and 3). XIS1 has a back-illuminated (BI) sensor, while XIS0, 2, and 3 
have front-illuminated (FI) sensors. The averaged pointing direction 
of the XIS is centered on (R.A., Dec.)=(\timeform{12h18m50.8s}, 
\timeform{+47D14'15.4''})\@. The XIS was operated in the normal 
clocking mode (8 s exposure per frame), with the standard 
5 $\times$ 5 and 3 $\times$ 3 editing mode. 
We processed the XIS data by the ``xispi'' and ``makepi'' ftools task 
and the CALDB files of 2008-01-31 version. 
Then, the XIS data were cleaned by assuming thresholds on the Earth elevation angle 
of $> 5^{\circ}$ and the Day-earth elevation angle of $> 20^{\circ}$. 
We also discarded data during the time since the spacecraft exit from 
the south Atlantic anomaly was unless than 436 sec. 
We created a light curve of each sensor over 0.5-2 keV binned 540 sec 
to exclude periods of an anomalous event rate greater or less 
than $\pm 3\sigma$ around the mean. 
After the above screening, the remaining good exposures were 99.9 and 99.8 ksec 
for FIs and BI, respectively.
Event screening with cut-off rigidity was not performed.
We also excluded the RAWY$<$119 region on the XIS2 and 3 due to 
an erroneous dark frame during this observation.

The spectral analysis was performed with HEAsoft version 6.5 and XSPEC 12.4.
In order to subtract the non-X-ray background (NXB), we employed 
the dark-Earth database by the ``xisnxbgen'' ftools task \citep{tawa_08}.
We generated two different Ancillary Response Files (ARFs) 
for the spectrum of each region, which assumed uniform 
sky emission and the observed XIS1 image by the ``xissimarfgen'' ftools 
task \citep{ishisaki_07}. We also included the effect of contaminations
on the optical blocking filters of the XIS in the ARFs.

\section{Analysis and Results}

\subsection{Strategy of Spectral Fitting}
\label{x-ray}

\begin{table*}
\caption{
The best-fit parameters of the apec component + pawer-law model.$^{\ast}$}
\label{par_bgd}
\begin{center}
\begin{tabular}{lcccc} \hline\hline
Parameters & & (i) & (ii)   \\ \hline
$\Gamma_{\rm CXB}$& & 1.38$\pm{0.05}$ & 1.37$\pm{0.05}$  \\ 
$Norm_{\rm CXB}\,^\dagger$& & 1.16$^{+0.05}_{-0.06}$ & 1.15$^{+0.05}_{-0.06}$  \\ 
$N_{\rm H}$ & (cm$^{-2}$) & 1.2$\times 10^{20}$ (fix) & 1.2$\times 10^{20}$ (fix) \\[1.0ex] 
$kT_{\rm MWH}$ & (keV) & 0.3 (fix) & 0.3 (fix)  \\ 
Abundance & (solar) & 1 (fix) & 1 (fix) \\ 
$Norm_{\rm MWH}\,^\ddagger$ & & 0.64$\pm{0.06}$ & 0.68$\pm{0.06}$ \\
 [1.0ex]
$kT_{\rm LHB}$ & (keV) & 0.1 (fix) & 0.1 (fix) \\ 
Abundance & (solar)& 1 (fix) & 1 (fix) \\ 
$Norm_{\rm LHB}\,^\dagger$ & & 1.74$^{+0.36}_{-0.37}$ & 1.80$^{+0.36}_{-0.37}$ \\[2.0ex] 
${\chi^2}$/d.o.f. & & 562/426 & 558/426 \\ \hline \hline
\end{tabular}
\end{center}
\parbox{\textwidth}{\footnotesize
\footnotemark[$\ast$]
The apec components for the spectra in the background region of 
NGC 4258 with absorbed or non-absorbed MWH component for the Galactic 
emission, and a power-law model for CXB. \\
\footnotemark[$\dagger$] 
Normalization of the power-law component
divided by the solid angle, $\Omega^{\makebox{\tiny\sc u}}$,
assumed in the uniform-sky ARF calculation (20$'$ radius),
in units of 10$^{-3}$ $\Omega^{\makebox{\tiny\sc u}}$ photons 
keV$^{-1}$ cm$^{-2}$ s$^{-1}$ arcmin$^{-2}$ at 1 keV.\\
\footnotemark[$\ddagger$]
Normalization of the apec components
divided by the solid angle same as the normalization of apec,
${\it Norm} = \int n_{\rm e} n_{\rm H} dV \,/\,
(4\pi\, (1+z)^2 D_{\rm A}^{\,2}) \,/\, \Omega^{\makebox{\tiny\sc u}}$
$\times 10^{-17}$ cm$^{-5}$ arcmin$^{-2}$, 
where $D_{\rm A}$ is the angular distance to the source.}

\end{table*}

Figure \ref{image} shows the observed X-ray 
contours in the 0.5--2 keV and 2--10 keV ranges, overlaid on the 
optical image from the Digitized Sky Survey (DSS). 
We extracted spectra from the following two regions. 

\begin{enumerate}
\item NGC 4258 components: an ellipse region, with semimajor and 
      semiminor axes of 8.3$'$ and 3.6$'$.

\item Background: the entire XIS field of view was used, after 
      excluding the above NGC 4258 component and three additional 
      1.5$'$ radius circular point source regions mentioned below.
\end{enumerate}

Three X-ray point-like sources were detected in 0.5--2 keV as shown
in figure \ref{image}. Sources \#1, 2, and 3 were identified as a 
quasar Q 1218+472 \citep{burbidge_95}, an Ultra Luminous X-ray 
source \citep{arp_04}, and an irregular galaxy UGC 7356 
\citep{thuan_79}, respectively. Although these sources were not 
so bright as to affect the spectra of NGC 4258, these were not negligible 
with respect to the estimation of the extra-galactic cosmic X-ray 
background (CXB) and the Galactic emissions.
In addition, three calibration sources, which have emission peaks 
at 5.9 keV, are located at the corners of the XIS. We included these regions for 
improvement in photon statistics because our target spectra are extracted only 
over the energy range below 5 keV.

The 0.5--2 keV contours of NGC 4258, which mainly represent 
the ISM emission, are clearly extended as compared to those 
in 2-10 keV, which are dominated by the LLAGN emission.
Therefore, the soft X-ray signal is considered to arise mainly 
from an extended hot halo. 
In fact, \citet{yang_07} showed that the 0.2--10 keV spectra observed with XMM-Newton 
were represented by the sum of two thermal components of 0.60 and 0.22 keV, 
plus the emission from LMXBs and the obscured LLAGN. 
The resultant parameters of these components imply that, even in the nuclear 
region within 3$'$, the diffuse thermal ISM emission dominate the total flux 
below $\sim2$ keV\@. Thus, we also analyzed the XIS spectra below 2 keV 
to investigate the ISM properties.

\subsection{Estimation of the background spectra}
\label{back}

In order to estimate the CXB and Galactic emission background components 
we first fitted the spectra of the background region. 
We assumed a power-law model for the CXB component, and a two 
temperature model for the Galaxy to represent emission from 
local hot bubble (LHB) and Milky Way halo (MWH).
We tested the following two models:
(i) phabs $\times$ power-law + apec$_{\rm MWH}$ + apec$_{\rm LHB}$, and 
(ii) phabs $\times$ (power-law + apec$_{\rm MWH}$) + apec$_{\rm LHB}$, 
where the apec models have a metal abundance fixed at 
the solar value and zero redshift. The absorption column density 
was also fixed to the Galactic value in the direction of NGC 4258, 
which is 1.2$\times10^{20}$ cm$^{-2}$. 
The difference between the two models is whether the MWH component is
absorbed by the Galactic column or not \citep{yamasaki_08}. 
We also fixed the temperature of the LHB and MWH emission 
to be 0.1 and 0.3 keV, respectively, after previous studies 
\citep{lumb_02,yamasaki_08,komiyama_09}. Due to poor photon statistics, 
leaving either temperature free did not provide useful constraints.
The spectra from the BI and FI CCDs were fitted simultaneously in 
the 0.5--5.0 keV range. 
The results of these fits are summarized in table \ref{par_bgd}. 
The normalization of the power-law model was found to be higher ($\sim$19\%) 
than the previously reported value (e.g., \cite{kushino_02}). 
This may be accounted for as due to the CXB brightness fluctuation, or 
to possible contamination by point sources in NGC 4258. 
Although the results do not differ significantly 
between the two models, model (ii) gave a marginally better fit statistic, 
${\chi^2}$/d.o.f. $=$ 558/426. 

In order to estimate the Galactic 
emission, we fitted the O\emissiontype{VII} and O\emissiontype{VIII} 
lines with Gaussian and a power-law in the 0.5--0.7 keV range.
The resultant intensities of the O\emissiontype{VII} and
O\emissiontype{VIII} without correcting for absorption 
are 3.8$\pm$1.1 and 1.1$\pm$0.9 photons cm$^{-2}$ s$^{-1}$ sr$^{-1}$, 
respectively. These are consistent with previously reported values
\citep{mccammon_02,sato_07a,yamasaki_08}.  
Hereafter, model (ii) is utilized as the background for 
the spectral fitting, unless otherwise stated.

\subsection{The galaxy spectra of NGC 4258}
\label{gal}

We fitted the galaxy spectra with the model: 
phabs$_{\rm G}\times$(vapec$_{\rm 1T, 2T, or 3T}$ + bremss) 
+ phabs$_{\rm AGN}\times$ power-law$_{\rm AGN}$ 
+ phabs$_{\rm G}\times$ (power-law$_{\rm CXB}$ + apec$_{\rm MWH}$) 
+ apec$_{\rm LHB}$, where the last term represents the 
background model to be explained later. 
In the model, phabs$_{\rm G}$ means the Galactic 
absorption in the direction of NGC 4258, fixed at 
$N_{\rm H} = 1.2\times10^{20}$ cm$^{-2}$. 
The term (phabs$_{\rm AGN}\times$ power-law$_{\rm AGN}$) shows 
the LLAGN contribution; its absorbing column was fixed at 
1.07$\times10^{20}$ cm$^{-2}$, and the power-law$_{\rm AGN}$ 
slope and normalization at $\Gamma=1.86$ and
$Norm=4.22\times10^{-3}$ photons keV$^{-1}$ cm$^{-2}$ s$^{-1}$ at 1 keV, 
both after \citet{yamada_09}. 
This formalism assumes that the absorption within NGC 4258
follows the same abundance ratios as the Galactic ones employed
in the phabs model. This assumption is justified {\em a posteriori}
by our result to be obtained below, that NGC 4258 and our Galaxy
have rather similar abundance patterns.
Although the intensity of the LLAGN component 
is much lower than that of the ISM components below 2 keV (subsection \ref{x-ray}), 
we take this term into account to decide accurately the emission 
of the Si-K$\alpha$ line around 1.8 keV where 
the LLAGN emission may affect the spectral fits.
The term (phabs$_{\rm G}\times$ 
(power-law$_{\rm CXB}$ + apec$_{\rm MWH}$) + apec$_{\rm LHB}$) 
represents the background component as examined in subsection \ref{back}.

The ISM emission of NGC 4258 was modeled with one, two, or
three-temperature models, as indicated by the subscripts 1T, 2T, or 3T, 
respectively, employing the vapec code \citep{smith_01}. 
The abundances of He, C, and N were fixed 
to the solar value. We also divided the other metals into 
five groups as O, Ne, (Mg \& Al), (Si, S, Ar, Ca), and (Fe \& Ni), 
based on the metal synthesis mechanism of SNe, and allowed 
them to vary.
The abundances were constrained to be common all temperature components. 
The bremss model, with $kT=10$ keV, represents the integrated LMXB component 
\citep{makishima_94,yamada_09}.
In order to constrain the background component contained in the above 
fitting model, we simultaneously fitted the source and background 
spectra, over the 0.5--2 keV and 0.5--5 keV regions, respectively. 
When fitting the background spectra, the normalizations of vapec, 
bremss, and power-law$_{\rm AGN}$ were all fixed to be 0. 

\begin{figure*}
\FigureFile(\textwidth,\textwidth){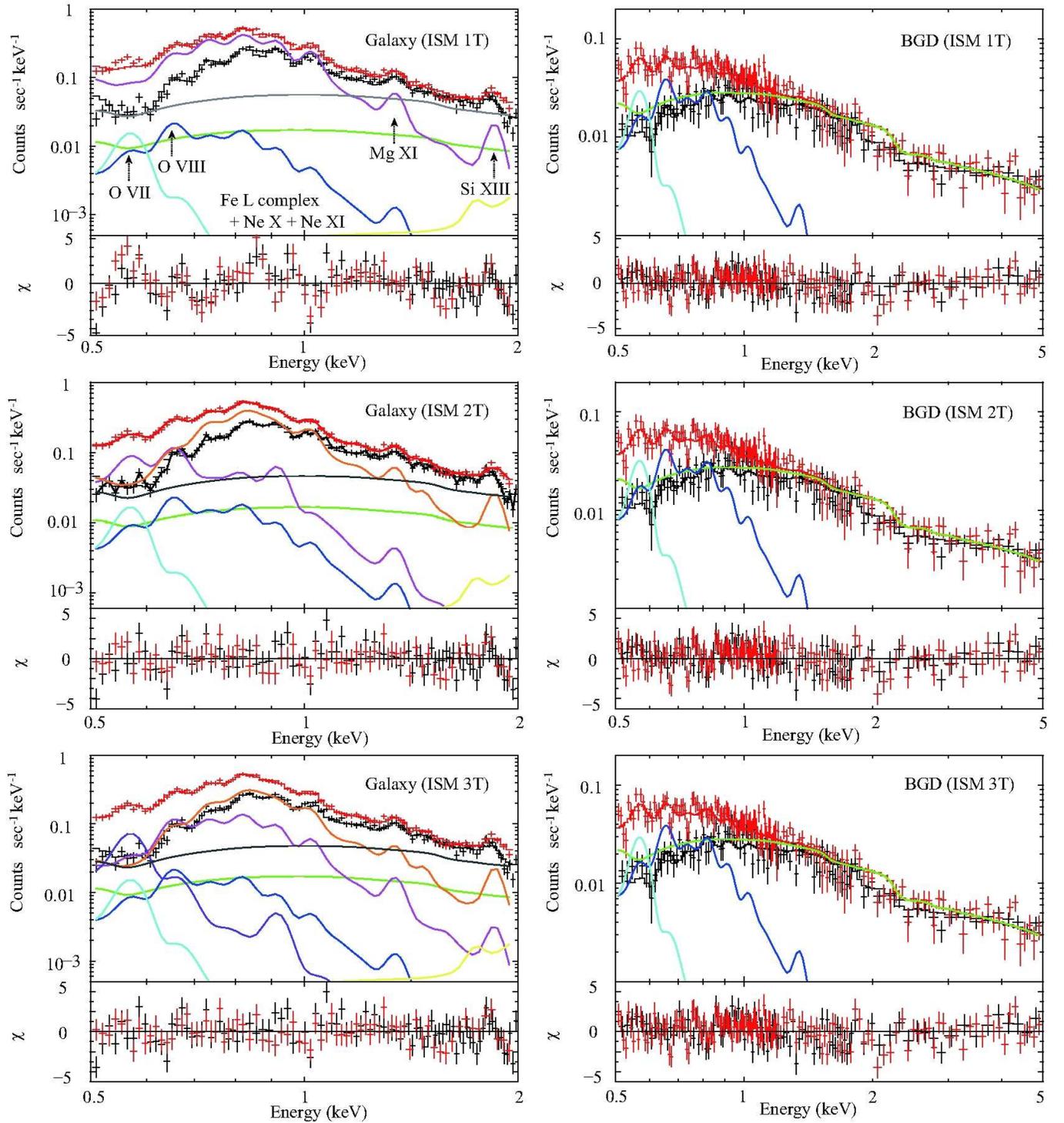}
\caption{
The NXB-subtracted XIS-FI (black) and XIS-BI (red) spectra 
of NGC 4258 (left column), and those of the background region, 
shown without removing the instrumental responses. 
Each spectrum from the top to bottom panel was fitted with a 
one, two, or three-temperature model for the ISM, respectively.
The black and red lines show the best-fit model for the FI and BI, 
respectively. For simplicity, only the model components for BI spectra 
are shown. The magenta, orange, and purple lines are the ISM components, 
blue and cyan are the Galactic background emission by apec$_{\rm MWH}$ 
and apec$_{\rm LHB}$, and green, gray, and yellow are the CXB, LMXB, and 
LLAGN components, respectively. 
The background components are common between the on-source and 
background spectra.
}
\label{spec_gal}
\end{figure*}

\begin{table*}
\caption{
Summary of the best-fit parameters of the NGC 4258 and background 
region.}
\label{par_gal}
\begin{center}
\begin{tabular}{lcccccc} \hline \hline
Parameters && 1T for ISM  & 2T for ISM & 2T for ISM& 3T for ISM \\ 
	& & & & (LHB and MWH free) & \\ \hline
$\Gamma_{\rm CXB}$ & & 1.42 & 1.41$\pm0.04$ & 1.31$^{+0.04}_{-0.06}$ & 1.41$\pm0.04$ \\ 
$Norm_{\rm CXB}\,^{\ast}$ & & 1.21 & 1.21$\pm0.04$ & 1.08$^{+0.03}_{-0.02}$ & 1.21$\pm0.04$ \\ 
$kT_{\rm MWH}$ & (keV) & 0.3 (fix) & 0.3 (fix) & 0.61$^{+0.57}_{-0.03}$ & 0.3 (fix) \\ 
$Norm_{\rm MWH}\,^{\dagger}$ & & 0.68 (fix) & 0.68 (fix) & 0.32$^{+0.03}_{-0.04}$ & 0.68 (fix) \\ 
$kT_{\rm LHB}$ & (keV) & 0.1 (fix) & 0.1 (fix) & 0.17$\pm0.01$ & 0.1 (fix) \\ 
$Norm_{\rm LHB}\,^{\dagger}$ & & 1.80 (fix) & 1.80 (fix) & 0.88$^{+0.07}_{-0.08}$ & 1.80 (fix) \\[2.0ex]
$kT_{\rm 1T}$ & (keV) & 0.38$\pm0.01$ & 0.23$\pm0.01$ & 0.24$\pm0.01$ & 0.13$^{+0.36}_{-0.06}$ \\ 
$Norm_{\rm 1T}\,^\dagger$ & & 2.18$\pm0.04$ & 0.68$^{+0.15}_{-0.19}$ & 0.72$^{+0.04}_{-0.05}$ & 0.41$^{+0.11}_{-0.02}$\\ 
$kT_{\rm 2T}$ & (keV) &-& 0.56$\pm0.01$ & 0.58$\pm0.01$ & 0.36$^{+0.06}_{-0.05}$ \\ 
$Norm_{\rm 2T}\,^\dagger$ & &-& 1.15$^{+0.06}_{-0.42}$ & 1.27$^{+0.26}_{-0.03}$ & 0.52$^{+0.03}_{-0.02}$ \\ 
$kT_{\rm 3T}$ & (keV) &-&-&-& 0.62$^{+0.06}_{-0.02}$\\ 
$Norm_{\rm 3T}\,^\dagger$ & &-&-&-& 0.68$^{+0.04}_{-0.02}$\\ 
O & (solar) & 0.36$^{+0.10}_{-0.07}$ & 0.56$^{+0.17}_{-0.08}$ & 0.53$^{+0.03}_{-0.04}$ & 0.91$^{+0.24}_{-0.32}$\\
Ne & (solar) & 1.20$^{+0.29}_{-0.20}$ & 1.28$^{+0.62}_{-0.20}$ & 1.12$^{+0.06}_{-0.07}$ & 1.43$^{+0.51}_{-0.46}$\\
Mg, Al & (solar) & 0.79$^{+0.21}_{-0.15}$ & 0.99$^{+0.47}_{-0.15}$ & 0.87$^{+0.13}_{-0.11}$ & 1.25$^{+0.42}_{-0.41}$\\
Si,S,Ar,Ca & (solar) & 1.59$^{+0.54}_{-0.39}$ & 1.11$^{+0.50}_{-0.18}$ & 0.99$^{+0.13}_{-0.11}$ & 1.42$^{+0.47}_{-0.45}$\\
Fe, Ni & (solar) & 0.49$^{+0.12}_{-0.08}$  & 0.64$^{+0.29}_{-0.09}$ & 0.57$^{+0.01}_{-0.02}$ & 0.84$^{+0.20}_{-0.27}$\\[2.0ex]
LMXB flux$^{\ddagger}$ &(erg cm$^{-2}$ s$^{-1}$) & 9.2 $\times 10^{-13}$ & 7.9 $\times 10^{-13}$ & 7.7 $\times 10^{-13}$ & 7.8 $\times 10^{-13}$ \\[2.0ex]
${\chi^2}$/d.o.f. & & 1333/740 & 993/738 & 946/734 & 982/736 \\ \hline \hline
\end{tabular}
\end{center}
\parbox{\textwidth}{\footnotesize
\footnotemark[$\ast$] 
Normalization of the power-law component
divided by the solid angle, $\Omega^{\makebox{\tiny\sc u}}$,
assumed in the uniform-sky ARF calculation (20$'$ radius),
in units of 10$^{-3}$ $\Omega^{\makebox{\tiny\sc u}}$ photons 
keV$^{-1}$ cm$^{-2}$ s$^{-1}$ arcmin$^{-2}$ at 1 keV.\\
\footnotemark[$\dagger$] 
Normalization of the vapec component scaled with a factor 
of {\sc source\_ratio\_reg} / {\sc area}, 
which is $Norm=\frac{\makebox{\sc source\_ratio\_reg}}
{\makebox{\sc area}} \int n_{\rm e} n_{\rm H} dV 
\,/\, [4\pi\, (1+z)^2 D_{\rm A}^{\,2}]$ $\times 10^{-17}$ cm$^{-5}$
 arcmin$^{-2}$, where $D_{\rm A}$ is the 
angular distance to the source.\\
\footnotemark[$\ddagger$]
Flux within the accumulated region between 0.5 and 2 keV.\\
} 
\end{table*}

The results of these fits are shown in figure \ref{spec_gal}, and the 
derived parameters are summarized in table \ref{par_gal}.
Thus, several emission lines are seen around 0.5--0.6 keV, 0.6--0.7 keV, 
$\sim$ 1.3 keV, and $\sim$ 1.8 keV, and are identified with those of 
O \emissiontype{VII}, O \emissiontype{VIII}, 
Mg \emissiontype{XI} and Si \emissiontype{XIII}, respectively. 
Furthermore, emission lines around 0.7--1 keV correspond to the 
Fe-L complex, as well as to K-lines from Ne \emissiontype{IX} and Ne \emissiontype{X}. 
The fits are not formally acceptable because of the big 
residuals around 2 keV owing to calibration issues of Si edges, 
which are known to be present in all the XIS CCDs \citep{koyama_07}.
However, these results are still useful to assess whether the temperatures 
and metal abundances in the ISM were reasonably determined or not.
The one-temperature model results in  
$kT=0.38$ keV\@. The fit statistics shown in table \ref{par_gal} 
clearly favor the two-temperature model, which 
gives the two temperatures as 0.23 and 0.59 keV\@, with 
corresponding 0.2--2.4 keV unabsorbed fluxes of 
1.0$\times10^{-12}$ and 2.5$\times10^{-12}$ erg cm$^{-2}$ s$^{-1}$, respectively. 
The 0.1--2.4 keV total unabsorbed ISM luminosity 
is 2.2$\times 10^{40}$ erg s$^{-1}$, which is 
consistent with that of the previous XMM-Newton results \citep{yang_07}.
The three-temperature model improved the fit statistics only slightly, and 
yielded the temperatures as 0.13, 0.36, and 0.62 keV\@. 
The implied abundance ratios did not
change, as shown in table \ref{par_gal}. 

\begin{figure*}
\FigureFile(\textwidth,\textwidth){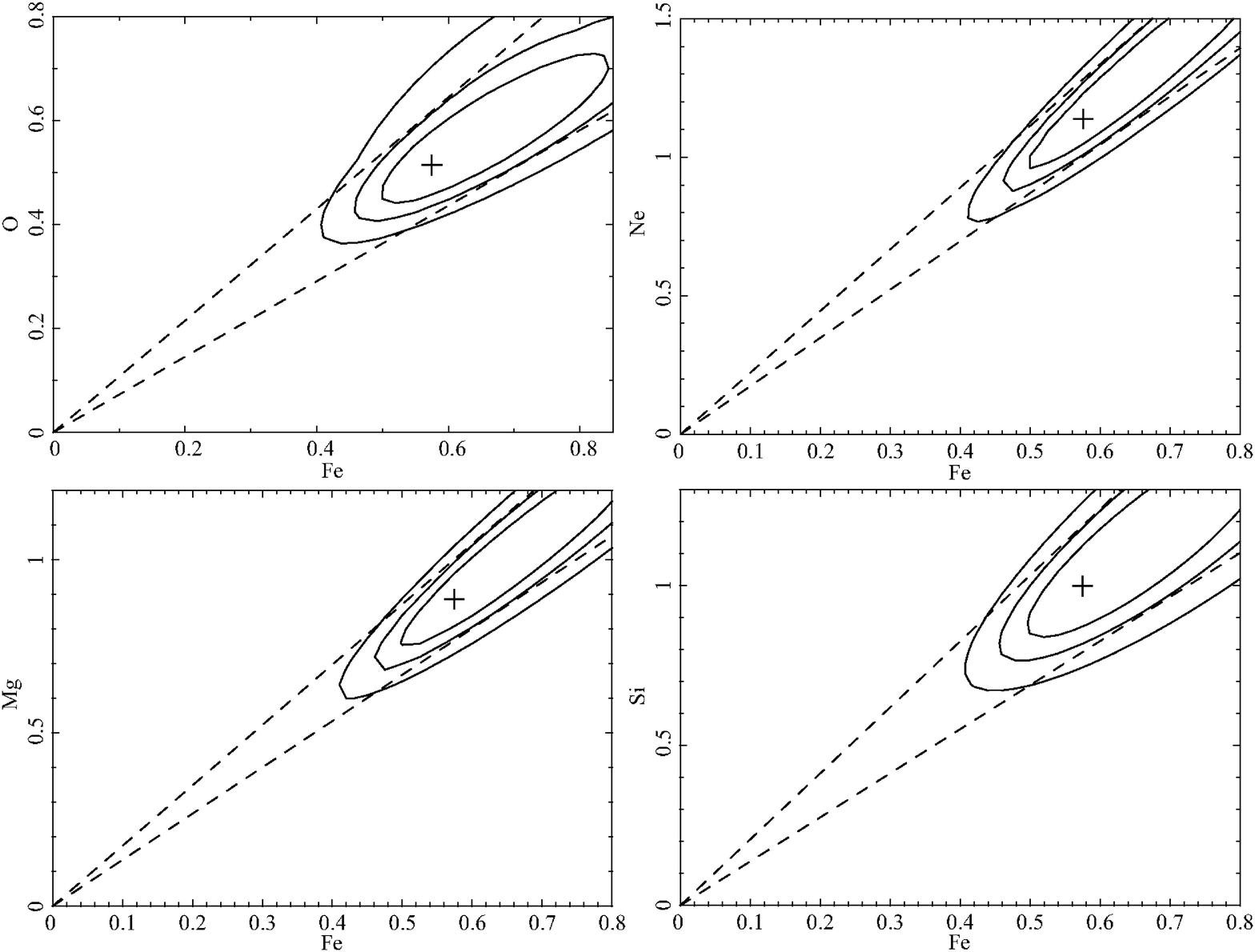}
\caption{
Plot of the confidence contours between the metal (O, Mg, S, Si) and Fe 
abundances, indicated by the two-temperature model for the ISM.
The three contours represent 68\%, 90\%, and 99\% confidence range 
from inner to outer, respectively.  
}\label{contour}
\end{figure*}

In order to examine abundance ratios rather than absolute values, 
we calculated confidence contours between the abundance of 
metals (O, Ne, Mg and Si) and that of Fe, 
using the two-temperature model for the ISM. 
The results are shown in figure \ref{contour}, 
where we also indicate 90\%-confidence abundance of these metals 
relative to Fe. 

\subsection{Systematic Uncertainties in the Abundance Ratios}

In order to investigate how these results are sensitive to the assumed 
Galactic model, we let the temperatures of 
the MWH and LHB vary freely. Although the temperatures of 
the Galactic components became 0.17 and 0.61 keV, the 
ISM abundance ratios were little affected. 
When we change the normalizations of Galactic 
apec$_{\rm MWH}$ and apec$_{\rm LHB}$ by $\pm10$\% and $\pm20$\%,
respectively, 
the metals to Fe abundance ratios of the ISM remained the same 
within $\sim5$\%. 
The systematic uncertainty 
due to the assumption of the OBF contaminant is less than 
the statistical error in the NGC 4258 region. 
Thus, the uncertainties in the background and the OBF contaminant
have little effect on the abundance ratios of the NGC 4258 ISM. 

Next, we investigate uncertainties in the absorption parameter. 
Although we have carried out the fits with the absorption fixed to the 
Galactic value, \citet{yang_07} have shown that absorption 
varies by a factor of a few, depending on location.
When we fit the spectra with the absorption allowed to vary freely, the resultant 
absorption is constrained with an upper limit of $N_{\rm H}$ of $4.7\times10^{20}$ cm$^{-2}$.
Thus, we confirm that the effect of the absorption is almost negligible 
because the resultant abundance change insignificantly. 

In order to examine the dependence on the LMXB component,
we used power-law model instead of the bremss model 
in \citet{makishima_94} and \citet{yamada_09}.
However, the abundance ratios to Fe did not change significantly either. 

The derived Ne to Fe abundance ratio is higher by a factor of $\sim2$ 
higher than the solar abundance. 
This is presumably because the Ne abundance is not reliably determined 
due to an overlap with the strong and complex Fe-L line emissions.

\citet{yang_07} have detected the hot spots of the jet located around 1$'$ 
from the nucleus in X-rays, and also found a non-thermal component besides 
that from the AGN within a \timeform{20''} region centered on the nucleus. 
In order to examine the contribution of the jet emission, 
we tried to investigate the spatial distribution of the metal abundances. 
Because the above physical scales are much smaller than the nominal 3$'$ 
point spread function of Suzaku \citep{serle_07}, we are only able to 
place crude constraints. 
We first extracted spectra from two separate regions for comparison: 
(i) a circular aperture of radius 3$'$ centered on the nucleus, and 
(ii) the remainder of the elliptical data-accumulation region from 
figure \ref{image}, excluding the central 3$'$.
We fitted the two spectra with the two-temperature model.
Unfortunately, poor photon statistics prevented the detection 
of any significant abundance differences between the two regions.
Thus we carried out a different test in which we added the following 
jet model from \citet{yang_07} to the two-temperature model
: phabs$_{\rm jet}\times$power-law$_{\rm jet}$.
Here, phabs$_{\rm jet}$, the intrinsic absorption, was fixed at 
3.4$\times10^{21}$ cm$^{-2}$, and the photon index of 
power-law$_{\rm jet}$ was fixed at 1.74 
\citep{yang_07}. 
Then, the 0.5--2 keV jet emission flux was rather tightly constrained 
to be $<$ 4.5$\times10^{-14}$ erg cm$^{-2}$ s$^{-1}$, at a confidence 
limit of 90\%.
Because the other parameters, including the metal abundances, 
did not change significantly, we conclude that our results are not affected 
by the jet emission.

\begin{figure*}
\FigureFile(\textwidth,\textwidth){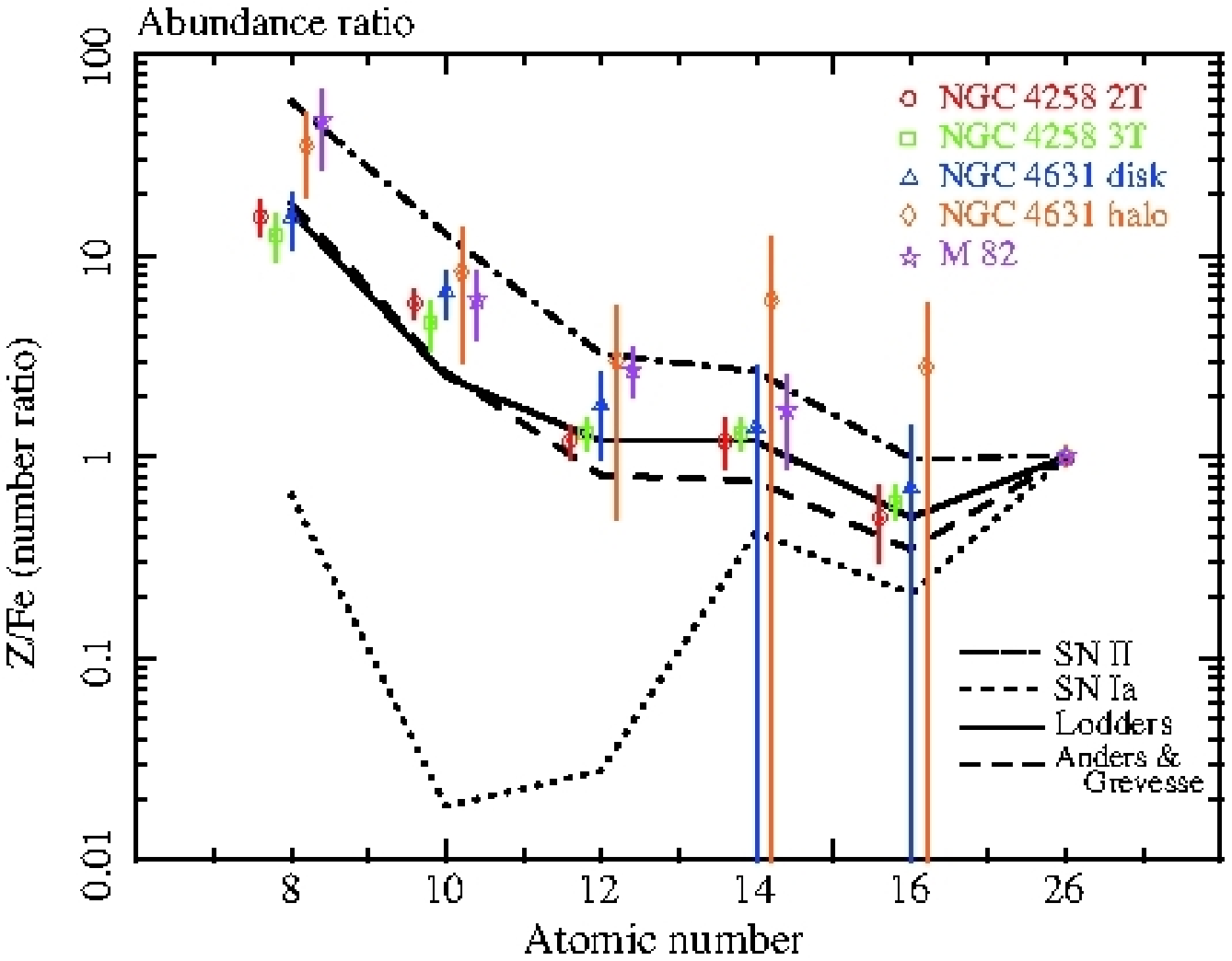}
\caption{
Number ratios of O, Ne, Mg, Si and S to Fe for the two and three-temperature
model of the ISM in NGC 4258. Solid and dashed lines indicate 
the number ratios of metals to Fe due to \citet{lodders_03} and 
\citet{anders_89}, respectively. 
Abundance patterns of NGC 4631 disk, halo \citep{yamasaki_08}, and M 82 ``cap'' 
\citep{tsuru_07} also are shown. 
Dot-Dashed and dotted lines represent the number ratios of metals to 
Fe for the SN II and SN Ia products \citep{iwamoto_99, nomoto_06}.
}\label{ratio}
\end{figure*}

\section{Discussion}\label{sec:discuss}

The present Suzaku observation has clearly revealed emission line features 
from the ISM in the spiral galaxy NGC 4258. 
We successfully measured the metal abundances of O, Ne, Mg, Si 
and Fe for the first time. 
Previous observations with 
XMM-Newton and Chandra with better angular resolutions 
successfully measured spatial temperature variations 
and the geometric structure of ``anomalous arms''. 
But, the metal abundances were not constrained due to insufficient photon 
statistics \citep{yang_07}. 
Figure \ref{ratio} shows our metal-to-Fe ratios, which were derived 
from the two-parameter confidence contours in figure \ref{contour}. 
The ISM emission is here modeled with two temperatures. 

In figure \ref{ratio}, the abundance pattern of NGC 4258 is compared with 
those indicated by the solar abundance table  of  \citet{anders_89} and 
the new solar abundance table of \citet{lodders_03}.
The O abundance in \citet{lodders_03}  was derived from solar 
photospheric lines, considering three-dimentional hydrostatic model
atmospheres and non-local thermodynamic equilibrium (Asplund 2005 
and references therein).
Thus, the abundance pattern of the ISM of NGC 4258 measured with Suzaku 
agrees better with that of \citet{lodders_03}, rather than that of \citet{anders_89}.

The calculated SN II and SN Ia yields are also plotted in figure \ref{ratio}.
Here, the SN II yields by \citet{nomoto_06} refer to an 
average over the Salpeter initial mass function of 
stellar masses from 10 to 50 $M_{\odot}$, with a progenitor 
metallicity of $Z=0.02$\@.
The SN Ia yields  were taken from the W7 model
\citet{iwamoto_99}.
The abundance pattern of the ISM of NGC 4258 is located between
those of SN II and SN Ia, and consistent with their mixture is. 

In order to investigate differences between spiral and 
starburst galaxies, we also plot in figure \ref{ratio} 
the abundance pattern of the hot ISM in 
disk and halo regions of the starburst galaxy NGC 4631 
 \citep{yamasaki_08}, and that of the ``cap'' region of the extreme 
starburst galaxy M 82 \citep{tsuru_07}.
The results of NGC 4631 disk are consistent with 
those of \citet{lodders_03} as is the case of NGC 4258, 
while the respective patterns of M 82 ``cap'' and  the halo region 
of NGC 4631 are close to those expected from SN II yields.

The SFR of NGC 4258 derived from its far infrared luminosity 
is low, $\sim$0.01 $M_{\odot}$ yr$^{-1}$ \citep{wu_06}, 
compared with those of NGC 4631 and M 82,
which are 2.99 and 9.39 $M_{\odot}$ yr$^{-1}$, respectively 
\citep{tullmann_06}.
The 0.5--2 keV luminosity of the LMXB component of NGC 4258, 
3.5 $\times$ 10$^{39}$ erg s$^{-1}$, is consistent with 
the mass vs. X-ray luminosity relation for spiral galaxies \citep{gilfanov_04}. 
Thus, we can regard NGC 4258 as a ``normal'' spiral galaxy. 
Therefore, we may conclude that solar abundance pattern are common in normal spiral 
galaxies, including NGC 4258 and our Galaxy.
The fact that the disk region of NGC 4631 also has a similar abundance pattern
to normal spirals, suggests that the 
metallicity of the ISM after its starburst era may look quite similar.
In contrast, in starburst galaxies, SN II products 
such as O effectively escape into the halo region as a result of the 
energetic explosions.
Thus, observations of abundance patterns such as ours 
play a key role in investigating the processes of galaxy evolution 
and enrichment of the intergalactic medium.

\bigskip
We thank the referee for providing valuable comments.
We gratefully acknowledge all members of the Suzaku hardware and software 
teams and the Science Working Group. 
KS acknowledges support by the Ministry of Education, Culture, Sports, 
Science and Technology of Japan, 
Grant-in-Aid for Scientific Research No. 21740134. 
PG acknowledges a RIKEN Foreign Postdoctoral Research Fellowship.

\end{document}